\documentclass[letter]{aa}
\usepackage{appendix}
\usepackage{graphicx}
\usepackage{array}
\usepackage{txfonts}
\usepackage{longtable,ltcaption}
\usepackage{lscape}
\usepackage{natbib}
\usepackage{lscape}
\usepackage{amsmath}
\usepackage{wasysym}
\usepackage{graphics}
\usepackage{txfonts}
\usepackage{color}
\usepackage{times}
\usepackage{parskip}
\usepackage{pdflscape}
\usepackage{geometry}
\usepackage{marginnote}
\usepackage{stfloats}
\usepackage{mathtools} 
\usepackage{nicefrac}
\usepackage{float}
\usepackage[normalem]{ulem}

\usepackage{fixmath}

\makeatother

\newfont{\nlx}{cmssdc10 scaled 900}
\newfont{\mfont}{cmssdc10 scaled 760}
\definecolor{myblue1}{rgb}{0.0,0.604,0.831} 
\definecolor{myblue2}{rgb}{0.0,0.49,0.6745}
\definecolor{myblue3}{rgb}{0.0156,0.4078,0.9921}
\definecolor{myblue4}{rgb}{0.0,0.44,0.87}
\definecolor{myred1}{rgb}{0.529,0.019,0.017}
\definecolor{mycyan}{rgb}{0.63921569,0.0,0.48235294}

\newcommand{\brem}[1]{\textcolor{black}{\nlx #1}}

\def\sigmastar{$\sigma_{\star}$}
\def\taustar{$\tau_{\star}$}

\def\fado{{\sc Fado}}

\def\ppxf{{\sc pPXF}}
\def\bayes{{\sc Bayes-LOSVD}}
\def\kinemetry{{\sc Kinemetry}}
\def\photutils{{\sc PhotUtils.isophote}}

\def\D4000{$D_{4000}$}

\newfont{\hvss}{cmssdc10 scaled 1540}
\def\?{{\bf\color{red}?}}
\begin{document} 
\title{Large-Scale Stellar Age-Velocity Spiral Pattern in NGC 4030}

   \author{
          Iris Breda \inst{\ref{UniVie}}
          \and
          Glenn van de Ven \inst{\ref{UniVie}}
          \and
          Sabine Thater \inst{\ref{UniVie}}
          \and
          J. Falc\'on-Barroso \inst{\ref{IAC}, \ref{IAC0}}
          \and
          Prashin Jethwa \inst{\ref{UniVie}}
          \and
          Dimitri A. Gadotti \inst{\ref{CEA}}
          \and
          Masato Onodera \inst{\ref{Tokyo}, \ref{Subaru}}
          \and
          Ismael Pessa \inst{\ref{AIP}}
          \and 
          Joop Schaye \inst{\ref{Leiden}}
          \and
          Gerhard Hensler \inst{\ref{UniVie}}
          \and
          Jarle Brinchmann \inst{\ref{UPorto}}
          \and
          Anja F.-Krause \inst{\ref{UniVie}} 
          \and
          Davor Krajnovi\'{c} \inst{\ref{AIP}}
          \and
          Bodo Ziegler \inst{\ref{UniVie}}}
          
\institute{Dep. of Astrophysics, University of Vienna, Türkenschanzstraße 17, 1180, Vienna, Austria \label{UniVie}
         \and
Instituto de Astrof\'isica de Canarias, Calle V\'ia L\'actea s/n, E-38205, La Laguna, Tenerife, Spain \label{IAC}
		\and
Dep. de Astrof\'isica, Universidad de La Laguna, Av. del Astrof\'isico Francisco S\'anchez s/n, E-38206, La Laguna, Tenerife, Spain \label{IAC0}
		 \and 
Centre for Extragalactic Astronomy, Department of Physics, Durham University, South Road, Durham DH1 3LE, UK \label{CEA}
         \and
Graduate Institute for Advanced Studies, SOKENDAI, 2-21-1 Osawa, Mitaka, Tokyo 181-8588, Japan \label{Tokyo}
         \and
Subaru Telescope, National Astronomical Observatory of Japan, 650 N Aohoku Pl, Hilo, HI96720 \label{Subaru}
         \and
Leibniz-Institut für Astrophysik Potsdam (AIP), An der Sternwarte 16, 14482 Potsdam, Germany \label{AIP}
         \and
Dep. de F\'isica e Astronomia, Faculdade de Ci\^encias, Rua do Campo Alegre, 4169-007 Porto, Portugal \label{UPorto}
  		 \and
Leiden Observatory, Leiden University, PO Box 9513, 2300 RA Leiden, the Netherlands \label{Leiden}		
\\
             \email{iris.breda@univie.ac.at}
             }

   \date{Received ???; accepted ???}

\abstract
{
The processes driving the formation and evolution of late-type galaxies (LTGs) continue to be a debated subject in extragalactic astronomy. Investigating stellar kinematics, especially when combined with age estimates, provides crucial insights into the formation and subsequent development of galactic discs. 
Post-processing of exceptionally high-quality Integral Field Spectroscopy (IFS) data of \object{NGC 4030} acquired with the Multi Unit Spectroscopic Explorer (MUSE), clearly reveals a striking grand design spiral pattern in the velocity dispersion map not previously detected in other galaxies. This pattern spatially correlates with HII regions, suggesting that stars currently being born exhibit lower velocity dispersion as compared to surrounding areas where star formation (SF) is less active. We examine the age-velocity relation (AVR) and propose that its configuration might be shaped by a combination of heating mechanisms, seemingly consistent with findings from recent high-resolution cosmological zoom-in simulations. The complex structure of the uncovered AVR of NGC 4030 support the hypothesis that stellar populations initially inherit the velocity dispersion $\sigma$ of the progenitor cold molecular gas, which depends on formation time and galactocentric distance, subsequently experiencing kinematic heating by cumulative gravitational interactions during their lifetime. 
While advancing our understanding of the AVR, these findings offer a new framework for investigating disk heating mechanisms, and their role in the evolution of galactic disks.
}

\maketitle

\parskip = \baselineskip

\section{Introduction \label{intro}}
\nolinenumbers

Observations of individual stars in the solar vicinity show a clear trend of increasing stellar velocity dispersion (\sigmastar) with stellar age (\taustar), demonstrating that older stars are typically kinematically hotter \citep[e.g.,][]{Cas11} whereas younger stars tend to be kinematically cooler \citep[e.g.,][]{Agu01}. This is known as the stellar age-velocity relation (AVR), a well-established phenomenon in Galactic dynamics, though its origin remains highly debated. The AVR is frequently attributed to secular kinematical heating mechanisms reflecting the cumulative effect of gravitational interactions by giant molecular clouds \citep{SpiSch51,SpiSch53,KokIda92}, stellar populations in the bar \citep{Deh00,MinFam10}, spiral arms \citep{BarWol67,GolTre80,DeS04}, halo black holes \citep{HanFly02}, MAssive Compact Halo Objects \citep[MACHOs,][]{LacOst85,BinTre08}, and minor mergers \citep{Qui93,Wal96,HuaCar97}. Other authors, however, have invoked an alternative explanation: spatially resolved observations of high-redshift galaxies have revealed that disk galaxies at these epochs tend to exhibit more turbulent interstellar media (ISMs), as compared to local galaxies \citep[e.g.,][]{Fos09,Gen11,Kas12,Sto16}. Therefore, by assuming that stars inherit the kinematic imprints of their progenitor gas cloud, the AVR might simply reflect the gradual settlement of the gas in the disk, i.e., older stellar populations, born from more turbulent ISMs, inherit higher \sigmastar\ as compared to younger stellar populations that formed when the ISM was less turbulent.

In our own galaxy, the AVR offers crucial insights into the processes shaping the dynamical evolution of the stellar disk. However, different studies report various configurations of the Milky Way's (MW) AVR. Frequently, the AVR is modelled as a smooth power law indicating continuous heating, offering theoretical predictions on the evolution of \sigmastar\ (both radial and vertical) over time (\sigmastar\ $\propto$ \taustar$^x$), with values of $x$ ranging from 0.2 to 0.6 \citep[see][for details]{KumBabSai17,FatSaj23}. In contrast, other authors argue that \sigmastar\ increases sharply for stars up to 5 Gyr old followed by a plateau, suggesting saturation of the heating mechanism \citep{CarSel85,Gom97}. In addition, a third model postulates that dispersion changes in discrete age groups (younger than 3 Gyr, 3-10 Gyr, and older than 10 Gyr) without a smooth age correlation, possibly due to events such as mergers affecting older stars \citep{Fre91,QuiGar01}.


There are several works attempting to reproduce the MW's AVR via 
models and/or simulations, employing a variety of theoretical frameworks and computational techniques.
While these have significantly advanced our knowledge of the mechanisms causing the observed AVR, the outcomes can differ substantially and may sometimes be contradictory. A recurrent result is that stellar velocity dispersion depends on the combination of the initial velocity dispersion of the gas, determined by the turbulence of the ISM, and the subsequent gradual heating due to interactions with the aforementioned perturbative agents.
For instance, by employing smoothed particle hydrodynamic simulations of self-gravitating multiphase gas disks in static disk-halo potentials, \cite{KumBabSai17} found that the evolution of the stellar velocity dispersion is determined by the initial velocity dispersion and the heating rate. A similar conclusion is reached by models based on observed star formation histories (SFHs) of Local Group galaxies (LG), such as the ones by \cite{Lea17} where they adopt an evolutionary formalism to describe the ISM velocity dispersion influenced by the galaxy's evolving gas fraction. Their results suggest that stars are born with a velocity dispersion similar to that of the gas at their formation, with dynamical heating occurring with galaxy mass-dependent efficiency. 
In contrast, alternative interpretations have been proposed, such as those by van \cite{DonAgeRen22}, who suggest that the orbital eccentricities in the MW disc may stem solely from the characteristics of the preceding ISM, and by \cite{AuBinSch16}, whose N-body simulations indicate that gradual heating from giant molecular clouds (GMCs) and spiral/bar interactions may fully account for the MW's AVR.



Interesting insights emerge from cosmological zoom-in simulations, offering a detailed exploration of the intricate interplay between gravitational forces, gas dynamics, and stellar evolution across cosmic scales.
In this context, \cite{MarMinFly14a,MarMinFly14b} simulate seven disk galaxies with stellar masses ranging from 4.3 to 12 $\times 10^{10}$ $\rm M_{\odot}$. They find that old stars exhibit high \sigmastar, as every galaxy experiences an initial phase of active mergers at high redshift, leading to the early formation of a thick stellar component characterized by a high velocity dispersion. 
For the quiescent galaxies, referring to those that do not undergo any major subsequent merger, they note a gradual rise in \sigmastar\ with age, suggesting that the slope of the AVR is not predetermined at birth but is a consequence of subsequent heating mechanisms. Moreover, they observe a significant sensitivity of \sigmastar\ to the merger history. Another noteworthy result is the one by \cite{Bird21}. By employing a high-resolution cosmological zoom-in simulation of a MW-mass disc galaxy (h277), these authors find excellent agreement between the present-day AVR of h277 and that measured in the solar vicinity. They discover that older stars are born kinematically hotter, supporting an `upside-down' formation scenario, and are further heated after birth. Furthermore, the disc grows `inside-out', as the angular momentum of accreted gas increases with decreasing redshift. Interestingly, although the vast majority of old stars (> 10 Gyr) are born in the interior of the disc, many can be currently found in the solar neighbourhood, as a result of radial migration. As the velocity dispersion inherited at birth ($\sigma_{\rm birth}$) increases with lookback time and the outer disc remains dynamically colder than the inner disc, younger stellar populations are born at larger galactocentric radius with lower mean $\sigma_{\rm birth}$. Thereafter, subsequent heating due to encounters with perturbative agents accounts for approximately half of the present-day stellar velocity dispersion, $\sigma_{\rm birth}$/$\sigma \sim$ 0.4 – 0.5, with older (younger) stars displaying the lowest (highest) ratio. They also find a consistent agreement between the AVR measured for stars in the solar vicinity and that across the entire disc.

Beyond the MW, the AVR has been studied in LG galaxies such as Andromeda \citep[M31,][]{Dor15} and other nearby systems \citep{Lea17}. In addition, \cite{Pessa23} have performed the first assessment of the AVR outside the LG: 
analysis of 19 PHANGS-MUSE galaxies reveals that while spiral arms and disks, associated with younger stellar populations, share similar \sigmastar,  galaxy centres and bars, typically populated by older stars, display higher \sigmastar\ populations. To correctly interpret these trends, the authors explore \sigmastar\ radial gradients for different stellar age bins, uncovering that younger stars predominantly display lower \sigmastar. No significant differences were found between spiral arms and disks, suggesting that spiral arms do not play a primary role in stellar kinematic heating. 

In this study, we interpret the stellar AVR of \object{NGC 4030} through a detailed analysis of high-resolution integral-field spectroscopic data, following the unprecedented discovery of striking grand-design spiral patterns in the \sigmastar\ map, extending across the entire galaxy\footnote{Interestingly, similar \sigmastar\ features have previously detected in the innermost region of \object{NGC1097} \citep[see Fig. 2 of][]{Gad20}.}.
This spiral galaxy, at a distance of 29.9 Mpc, is morphologically classified as SA(s)bc, has an estimated total stellar mass of M$_{\star}$ = 1.5 $\times$ 10$^{11}$ M$_{\odot}$, an estimated star-formation rate of 11.08 M$_{\odot}$/yr, and an effective radius R$_{\rm eff}$ of 31.8'' $\sim$ 4.6 kpc \citep{Err19}.

\section{Methodology \label{meth}}

The spectroscopic data here analysed consists of Integral Field Unit (IFU) data observed with the Multi Unit Spectroscopic Explorer \citep[MUSE, ][]{MUSE} as part of the MUSE Atlas of Disks (MAD) Survey \citep{Err19}. The data-cube, which probes the innermost part of the galaxy covering approximately its effective radius, was analysed utilizing the pipeline comprehensively described in Breda et al., 2024b (in prep.). The data processing is initiated by correcting for Galactic extinction and de-redshifting, followed by Voronoi binning \citep[adopting the method by][]{CapCop03}, targeting at a signal-to-noise ratio (SNR) of 170 in the (emission-line-free pseudo-continuum) spectral range between 6390 and 6490 \AA. Stellar ages  were derived via spectral synthesis using \fado\ \citep{FADO}, within the spectral range between 4730 and 8740 \AA, employing the simple stellar population (SSP) library based on the \cite{BruCha03} models. This stellar library comprises 152 SSPs for 38 ages between 1 Myr and 13 Gyr for four stellar metallicities (0.05, 0.2, 0.4 and 1.0 $Z_{\odot}$), referring to a Salpeter initial mass function \citep{SalIMF} and Padova 2000 tracks \citep{Gir00}. \fado\ results indicate that the mean nebular extinction across the galaxy is A$_{\rm V}$ $\sim$1.6 mag.

Stellar kinematic extraction was carried out using both \ppxf\ \citep{CapEms04,Cap17}, fitting 6 and 4 Gaussian-Hermite moments (see Fig. \ref{app1}) with two sets of templates, the Indo-US Library of Coudé Feed Stellar Spectra \citep[IUS,][]{Val04} and MILES \citep[][]{MILES}, as well as \bayes\ \citep{FalMar21}, within the full spectral range and using SP (no regularization) by default, and RW (random-walk) if SP fails to fit the observed spectrum. Considering that star formation (SF) can lead to significant contamination of the stellar spectra by the nebular continuum, diluting absorption features which are vital for accurate stellar kinematic measurements, the spectra were decontaminated prior of fitting. Such was performed by assessing and subtracting the nebular continuum contribution, ensuring that only stellar emission (and emission lines, which will be masked) remain\footnote{For completeness, stellar kinematic maps obtained without subtracting the nebular continuum contribution are supplemented in the Appendix.}. Modelling of the nebular continuum was performed by following the technique from \cite{Pap98}, with adjustment of using the observed H$\alpha$ flux instead of the EW(H$\alpha$). Further details are provided in the appendix.

The underlying contribution of the main body of several physical properties was modelled with \kinemetry\ \citep{kinemetry} and subsequently subtracted (see Figs. \ref{app4}, \ref{app6}, \ref{app7}, \ref{app8}). This algorithm conducts harmonic expansion of 2D maps of observed moments, including surface brightness, mean velocity, and velocity dispersion, along the best-fitting ellipses, and can be employed in various other datasets which retain an underlying symmetrical structure. This step was instrumental, as it allowed to tentatively account for elementary substructures of various maps (e.g., the stellar kinematic component associated with the galaxy's gravitational potential). Retrieved residuals were analysed within and beyond the more prominent spiral arms, which were manually delineated upon the Hubble Space Telescope (HST) residual map. Radial profiles were determined using the python package \photutils\ \citep{Bra21}, representing the azimuthally-averaged flux in elliptical apertures (see Fig.~\ref{app2a}) as a function of radius. Uncertainties were derived by performing Monte Carlo (MC) simulations for 100 realizations, both for spectral synthesis 	and stellar kinematic extraction with \ppxf. In both cases, MC simulations were performed by adding random noise to the original spectrum, drawn from a normal distribution and scaled to match the typical residuals between the observed and the best-fit model. Finally, the AVR across the entire galaxy is examined, and we investigate how the mean \sigmastar\ and the luminosity-weighted stellar age, $\tau_{\star, \rm L}$, vary by selecting spaxels with progressively higher EWH($\alpha$).

\section{Results \label{res}}

In this section, we provide an overview of the main results obtained after processing the MUSE data-cube of \object{NGC 4030} with our pipeline (Breda et al. 2024b, in prep), and ensuing modelling and subtraction of the baseline contribution of maps of several physical properties, by means of \kinemetry.  In this fashion, maps of stellar age, stellar velocity dispersion, H$\alpha$, and EW(H$\alpha$) were retrieved, and radial profiles amidst and beyond the delineated spiral arms were computed. To provide a comparative reference, radial profiles of the entire map are additionally included.

\begin{figure*}[b]
\centering
\includegraphics[width=1\linewidth]{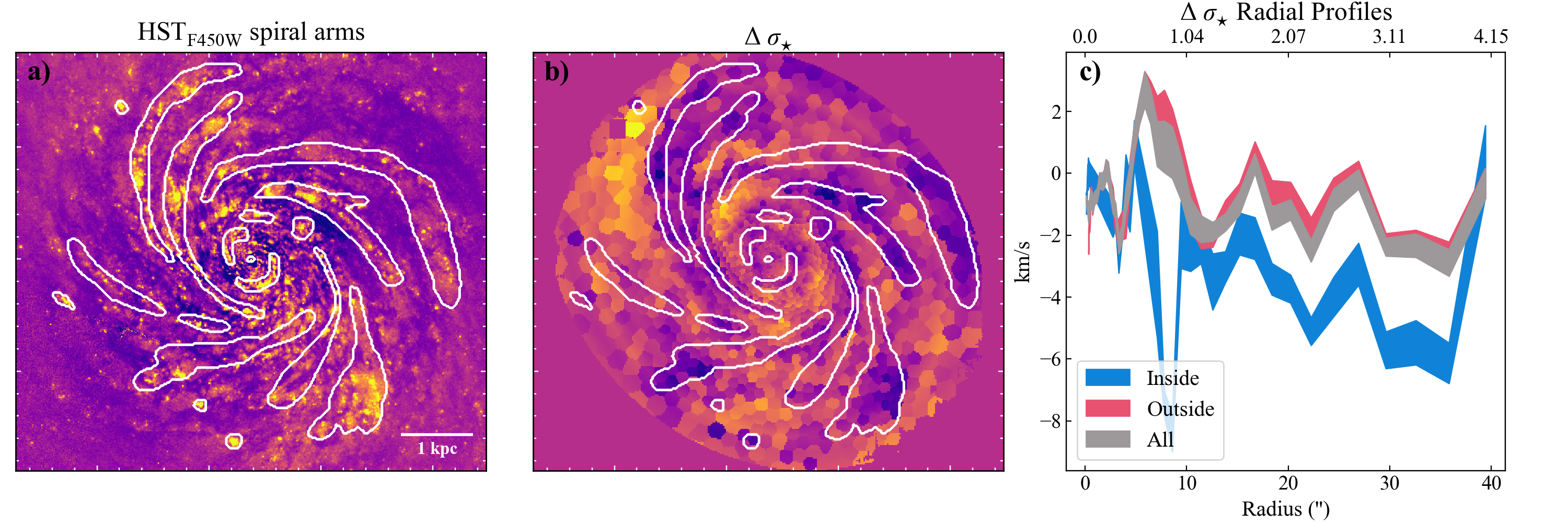}
\caption{Panel \brem{a}) Illustration of the selected regions highlighting the spiral arms of \object{NGC 4030}, having as reference the HST photometric frame in the filter F450W, after subtracting the contribution of the underlying main body with \kinemetry. Panel \brem{b}) Stellar velocity dispersion residual map ($\rm \Delta$\sigmastar) of \object{NGC 4030} with the spiral arms overlaid. Panel  \brem{c}) Radial profiles of $\rm \Delta$\sigmastar inside (blue line) and outside (red line) the delineated spiral arms, along with the overall profile for the entire map (grey line). The upper horizontal axis displays the distance from the center in kpc.}
\label{plot1}
\end{figure*}

Panel~\brem{a} of Fig.~\ref{plot1} displays our selection of the main spiral arms of \object{NGC 4030}, identified using the baseline-subtracted HST F450W photometric frame, i.e, after subtracting the best-fitting \kinemetry\ model. This selection will serve as a reference throughout this study, thereby defining the regions of interest to be contrasted with the remaining regions of the galaxy. Inspection of panel~\brem{b}, displaying the baseline-subtracted \sigmastar\ map ($\rm \Delta$\sigmastar), confirms that these regions of interest align with areas where \sigmastar\ decreases. The radial profiles presented in panel~\brem{c} reveal that within these regions, \sigmastar\ tends to be systematically lower by 2 – 8 km/s as compared to the surrounding areas. To note that, with a median uncertainty of 1.7 km/s in \sigmastar, $\rm \Delta$\sigmastar\ falls within the precision required to capture the observed variations. As expected, the radial profile of the entire map is consistent with the profile evaluated outside the delineated spiral arms regions, showing however lower values, due to the inclusion of the inner regions where \sigmastar\ is systematically lower. Interestingly, the spiral pattern is also clearly evident in the baseline-subtracted mean velocity map, which can be appreciated in the right panel of Fig.~\ref{app6}, resembling the phase-spirals observed in the MW \citep[e.g.,][]{MWS}.

\begin{figure*}
\centering
\includegraphics[width=1\linewidth]{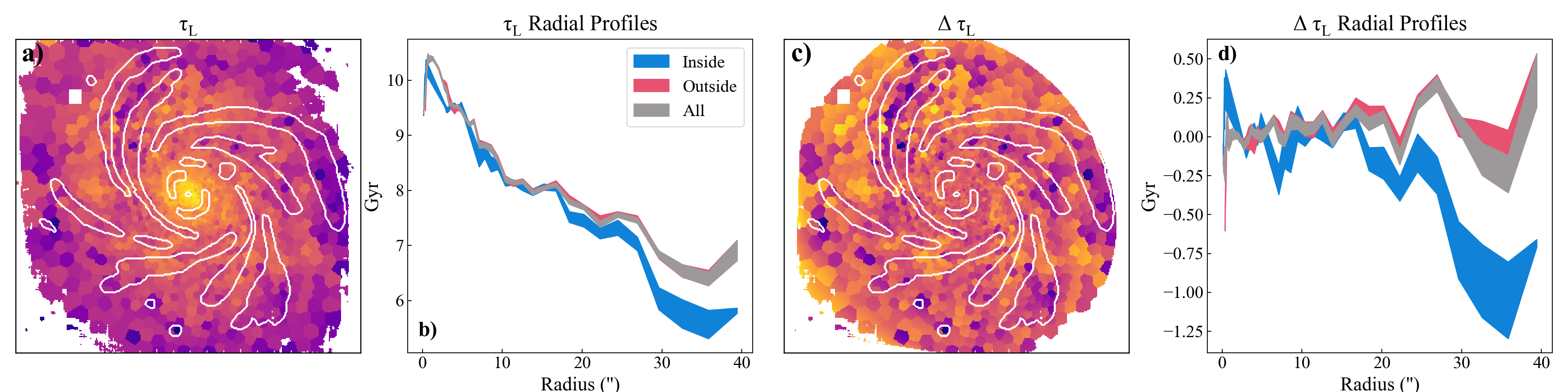}
\caption{Panels \brem{a}) \& \brem{b}) Luminosity-weighted stellar age assessed by \fado, followed by the respective radial profiles, inside (blue) and outside (red) the delineated spiral arms, with grey denoting the radial profile across the entire galaxy. Panels \brem{c}) \& \brem{d}) Comparable to panels \brem{a}) \& \brem{b}) but displaying instead the baseline-subtracted (i.e., after subtracting the \kinemetry\ best-fitting model) luminosity-weighted age map of \object{NGC 4030}.}
\label{plot2}
\end{figure*}

Subsequently, we aim at exploring how \sigmastar\ fluctuations correlate with the ages of the stellar populations and SF activity within these regions. Panel~\brem{b} of Fig.~\ref{plot2} displays the baseline-subtracted luminosity-weighted stellar age of \object{NGC 4030}. Similarly to other quantities under study, this map was obtained using \kinemetry\ to model and subtract the underlying structure of the 2D stellar age distribution derived by \fado. The radial profiles displayed in panel~\brem{c} confirm that the regions of interest are generally younger (0.25 – 1.25 Gyr, with a median uncertainty of 0.07 Gyr and maximum of 0.25 Gr) compared to the surrounding areas, particularly towards larger radii. Complementary, H$\alpha$ and EW(H$\alpha$) maps are shown in Fig.~\ref{plot3}, panels~\brem{a} and \brem{c}, respectively. The residual H$\alpha$ map, $\rm \Delta$H$\alpha$, was obtained by modelling and subtracting the best-fitting underlying main body from the H$\alpha$ map assessed by \fado\ using \kinemetry, while the EW(H$\alpha$) map derives directly from \fado. These maps clearly indicate that both $\rm \Delta$H$\alpha$ and EW(H$\alpha$) are significantly enhanced within the regions of interest compared to the surrounding areas, suggesting increased star formation activity. The combined evidence supports a coherent scenario: stellar populations currently forming in the spiral arms exhibit systematically lower \sigmastar\ compared to older stellar populations, which tend to be kinematically hotter. This finding aligns with the picture where stars are currently being born from cold gas, inheriting its low $\sigma_{\rm birth}$.

\begin{figure*}[!htbp]
\centering
\includegraphics[width=1\linewidth]{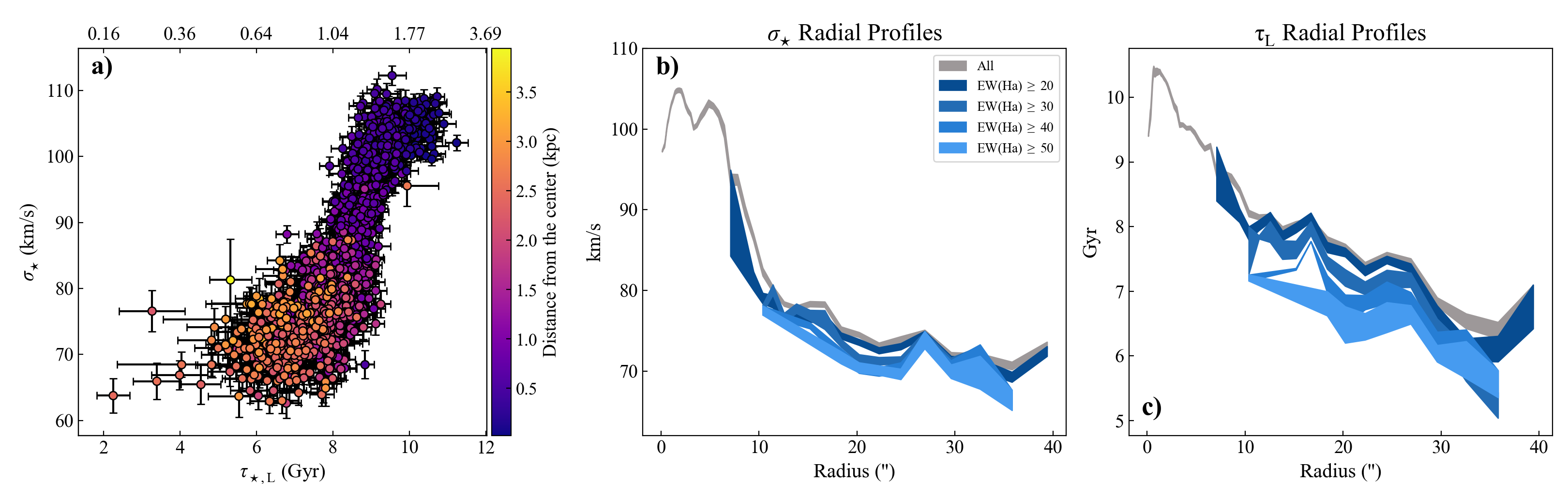}
\caption{Panel \brem{a}) Age-Velocity relation (AVR) of \object{NGC 4030} accounting for all bins displaying luminosity-weighted ages obtained with \fado. Values of the redshift at each $\tau_{\star, \rm L}$ are provided along the upper horizontal axis. Panel \brem{b}) Radial profiles of the observed \sigmastar\ for the whole map (grey) and by selecting spaxels that show increasing levels of EW(H$\alpha$) (in blue, with lighter shades corresponding to higher EW(H$\alpha$) levels). Panel \brem{c}) Similar to panel \brem{a}, illustrating the radial distribution of the luminosity-weighted stellar age.}
\label{plot4}
\end{figure*}

Panel \brem{a} of Fig.~\ref{plot4} displays the AVR for \object{NGC 4030} with luminosity-weighted stellar ages assessed by \fado, revealing an S-shaped pattern. It can be appreciated that, overall, younger stellar populations forming at larger galactocentric radii generally retain lower \sigmastar\ values, in contrast to the kinematically hotter, older populations in the galactic center and surrounding areas. Interestingly, stellar populations that originated around 7 – 8 Gyr ago exhibit a wide range of \sigmastar\ values that are strongly dependent on their galactocentric radius, suggesting a rapid formation phase across the sampled galactic region at this epoch. Nevertheless, the interpretation of the AVR is not straightforward. Speculatively, the complex structure of the AVR might indicate that several factors are at play, such as a combination of inherited \sigmastar\ at birth along with continuous kinematical heating by cumulative perturbations. We tend to discard major mergers as potential cause for the AVR, considering the galaxy's undisturbed morphology, although the incidence of early minor mergers cannot be excluded. Finally, to ensure that our findings are not driven by our customized selection delineating the spiral arms, nor by the subtraction of the underlying contribution from the various maps by means of \kinemetry, panels \brem{b} and \brem{c} display several \sigmastar\ and $\tau_{\star, \rm L}$ radial profiles, comparing their distributions across the entire galaxy with those derived from spaxels exhibiting progressively higher levels of EW(H$\alpha$). It is evident that \sigmastar\ in \object{NGC 4030} gradually decreases as spaxels with higher EW(H$\alpha$) levels are examined, revealing an anti-correlation between these physical properties. Unsurprisingly, a similar behaviour is documented for the luminosity-weighted stellar age, $\tau_{\star, \rm L}$.

\section{Conclusions}\label{conc}

Kinematic analysis of \object{NGC 4030} has revealed a grand design spiral pattern in the stellar velocity dispersion map which spatially correlates with HII regions. 
This outcome is independent of the empirically selected regions outlining the spiral arms, as higher EW(H$\alpha$) spaxels show lower \sigmastar\ and $\tau_{\star, \rm L}$. The presence of younger stellar populations and lower \sigmastar\ in these areas points to star formation from cold gas, where new stars inherit the gas's low $\sigma_{\rm birth}$. Comparably to our own MW and other galaxies, \object{NGC 4030}'s AVR indicates that younger (older) stellar populations generally exhibit lower (higher) \sigmastar.


Although the precise dynamical mechanisms driving the curved shape of  \object{NGC 4030}'s AVR remain challenging to determine, comparison with simulations might provide valuable insights. Admittedly, \object{NGC 4030} is significantly more massive than the MW (M$_{\rm \star, MW}\simeq0.6\times10^{11}$ M$_{\odot}$, \citealt{Lic15}, and M$_{\rm \star, NGC4030}\simeq1.5$ $\times$ \!$10^{11}$ M$_{\odot}$, \citealt{Err19}). Consequently, \object{NGC 4030} is also more massive than the simulated galaxy h277 from \cite{Bird21}. Nevertheless, the simulated AVR is notably steep for ages similar to those observed in our study (see their Figs. 1 \& 5), which is seemingly compatible with \object{NGC 4030}'s AVR. However, fundamental methodological differences obstruct direct comparison: while \cite{Bird21} provides the velocity dispersion and ages of individual stars, our analysis renders a line-of-sight velocity distribution (LOSVD) as well as luminosity-weighted ages across numerous stars, substantially enhancing our \sigmastar\ estimates and diluting our age estimates. Accordingly, we can only speculate that the apparent resemblance between the AVR from h277 and our galaxy might suggest that similar mechanisms are operating. Specifically, the presently observed \sigmastar\ may reflect both the primordial $\sigma_{\rm birth}$ from the progenitor gas, as well as gradual heating from cumulative interactions with perturbative agents during the stars' lifetime. Recalling the findings in h277, at higher redshift there is a greater fraction of available gas in the disk, increasing gravitational turbulence and resulting in stars born during these periods inheriting a higher $\sigma_{\rm birth}$. As turbulence is strong, favourable conditions for star formation are primarily attained towards the center and galactic plane, where stellar populations emerge first. Over time, as the disk settles, stars form at larger radii and latitudes from colder gas, inheriting a lower $\sigma_{\rm birth}$. Subsequent encounters with perturbing elements such as molecular clouds concentrated in the mid-plane, significantly contribute to the currently observed \sigmastar, promoting radial mixing and migration towards larger radii and higher latitudes. Concerning the study conducted by \cite{MarMinFly14b}, the majority of the simulated AVRs, particularly those without subsequent merger history, exhibit a similar pattern to the AVR of \object{NGC 4030} and h277. Interestingly, the AVR that most closely resembles our findings (after adjusting \sigmastar\ by summing a constant value) is associated with g106, the largest simulated galaxy with a 1:5 merger occurrence at $\tau$ about 7 Gyr in the past.


In conclusion, the seemingly corresponding structure between the simulated AVRs and that of \object{NGC 4030}, namely the S-shape for $\tau_{\star, \rm L} \gtrsim$ 4 Gyrs, might suggest that analogous physical mechanisms are at play. Within this framework, the AVR initially arises from the formation of stellar populations at different epochs, each originating from gas at varying temperatures/turbulence levels, thus retaining the imprint of their original $\sigma_{\rm birth}$. Subsequently, cumulative interactions with perturbative agents gradually increase the \sigmastar\ over the stars' lifetime. While other mechanisms, like past minor mergers, cannot be discarded, this scenario offers a fair explanation for our observations. 

The significant discovery of complex patterns in the \sigmastar\ map, tracing spiral arms and HII regions, offers new opportunities for comparative studies of disk heating mechanisms, and advances our understanding of galaxy formation by providing a rich dataset for benchmarking against simulations.

\begin{acknowledgements}
I.B. has received funding from the European Union's Horizon 2020 research and innovation programme under the Marie Sklodowska-Curie Grant agreement ID n.º 101059532.
I.P. acknowledges funding by the European Research Council through ERC-AdG SPECMAP-CGM, GA 101020943.
D.A.G. is supported by STFC grants ST/T000244/1 and ST/X001075/1.
J.F-B acknowledges support from the PID2022-140869NB-I00 grant from the Spanish Ministry of Science and Innovation.
\end{acknowledgements}


\appendix
\renewcommand\chaptername{Appendix}

\onecolumn

\begin{appendices}

\section{Supplementary maps for NGC 4030}\label{app}

\begin{figure}[h]
\centering
\includegraphics[width=0.9\linewidth]{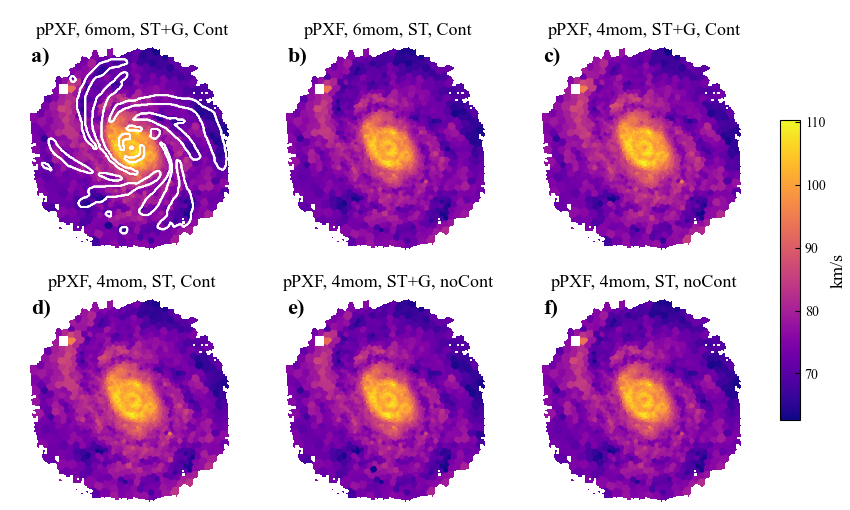}
\caption{\sigmastar\ maps of \object{NGC 4030} obtained with \ppxf\ with IUS templates by fitting -- panel \brem{a)} 6 kinematic moments, stellar absorption features and emission-lines simultaneously, on nebular-continuum subtracted spectra; panel \brem{b)} 6 kinematic moments, stellar absorption features while masking emission-lines, on nebular-continuum subtracted spectra; panel \brem{c)} 4 kinematic moments, stellar absorption features and emission-lines simultaneously, on nebular-continuum subtracted spectra; panel \brem{d)} 4 kinematic moments, stellar absorption features while masking emission-lines, on nebular-continuum subtracted spectra; panel \brem{e)} 4 kinematic moments,  stellar absorption features and emission-lines simultaneously, without subtracting the nebular-continuum; panel \brem{f)} 4 kinematic moments, stellar absorption features while masking emission-lines, without subtracting the nebular-continuum. Results utilising the MILES templates are equivalent. The \sigmastar\ spiral features are evident in all cases.}
\label{app1}
\end{figure}

In this work, contribution of the nebular continuum is modelled and subtracted prior to fitting with \ppxf. However, typical star-forming galaxies such as \object{NGC 4030}, with EWH($\alpha$) < 200 \AA, display week  nebular emission, and its correction has a small effect, which can be appreciated by comparing the the upper with the lower panels of Fig.~\ref{app1}). The methodology adopted for the decontamination of the spectra is provided bellow:\\
\brem{a}) H$\alpha$ and H$\beta$ fluxes are measured, allowing for the determination of intrinsic V-band extinction via the Balmer decrement;\\
\brem{b}) using a nebular continuum SED template as input, corresponding to standard conditions (electron temperature of 10.000 K and electron density of 100 cm$^{3}$), and computed following standard prescriptions \citep[e.g.,][]{Kru95,Fio97}:\\
\brem{b1}) the nebular continuum SED template is subjected to intrinsic extinction (based on the observed H$\alpha$/H$\beta$ ratio);\\
\brem{b2}) the modelled nebular continuum, to be subtracted to the observed spectra, is finally obtained by scaling the extinguished nebular continuum SED to the observed (therefore also extinguished) H$\alpha$ flux.

\begin{figure}[h!]
\centering
\includegraphics[width=0.9\linewidth]{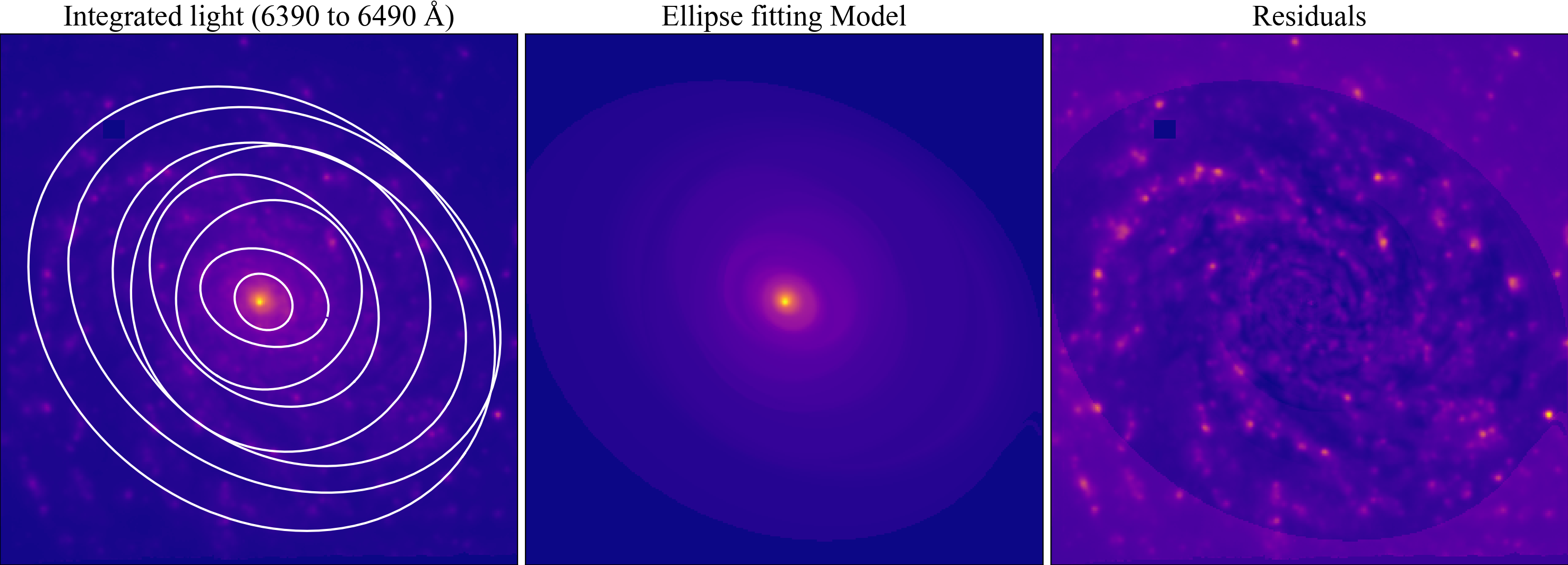}
\caption{Fitting elliptical isophotes to the MUSE integrated light within 6390 – 6490 \AA. Best-fitting ellipses are overlaid, shown alongside the model and residuals. These ellipses define the geometry used for extracting the radial profiles of the varied physical quantities assessed in this study.}
\label{app2a}
\end{figure}

\begin{figure}[h]
\centering
\includegraphics[width=1\linewidth]{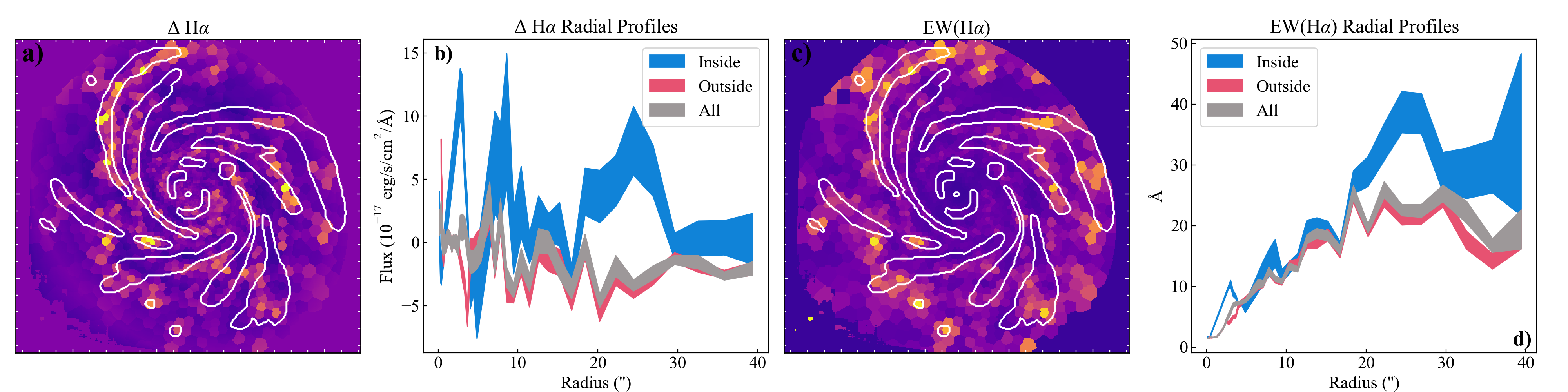}
\caption{Panel \brem{a} presents the baseline-subtracted H$\alpha$ map of \object{NGC 4030}, with the corresponding radial profiles displayed in panel \brem{b}. Panel \brem{c} shows the EW(H$\alpha$) map derived from \fado, i.e., without subtraction of the best-fitting underlying model, in contrast to previous maps. Panel \brem{d} showcases the radial profiles of EW(H$\alpha$). Consistent with previous figures, the spiral arms are superimposed in the maps, and the radial profiles evaluate regions inside and outside the spiral arms, as well as across the entire map.}
\label{plot3}
\end{figure}

\begin{figure}[h]
\centering
\includegraphics[width=1\linewidth]{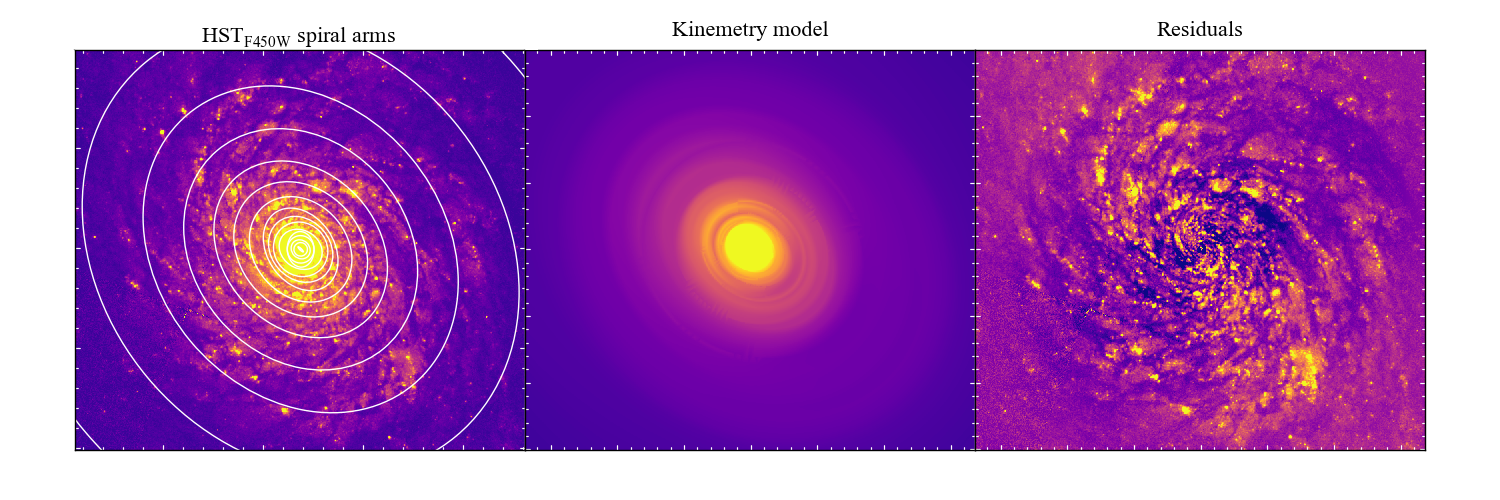}
\caption{Modelling and subtraction of the underlying substructure of HST F450W photometry image with \kinemetry.}
\label{app4}
\end{figure}


\begin{figure}[h!]
\centering
\includegraphics[width=1\linewidth]{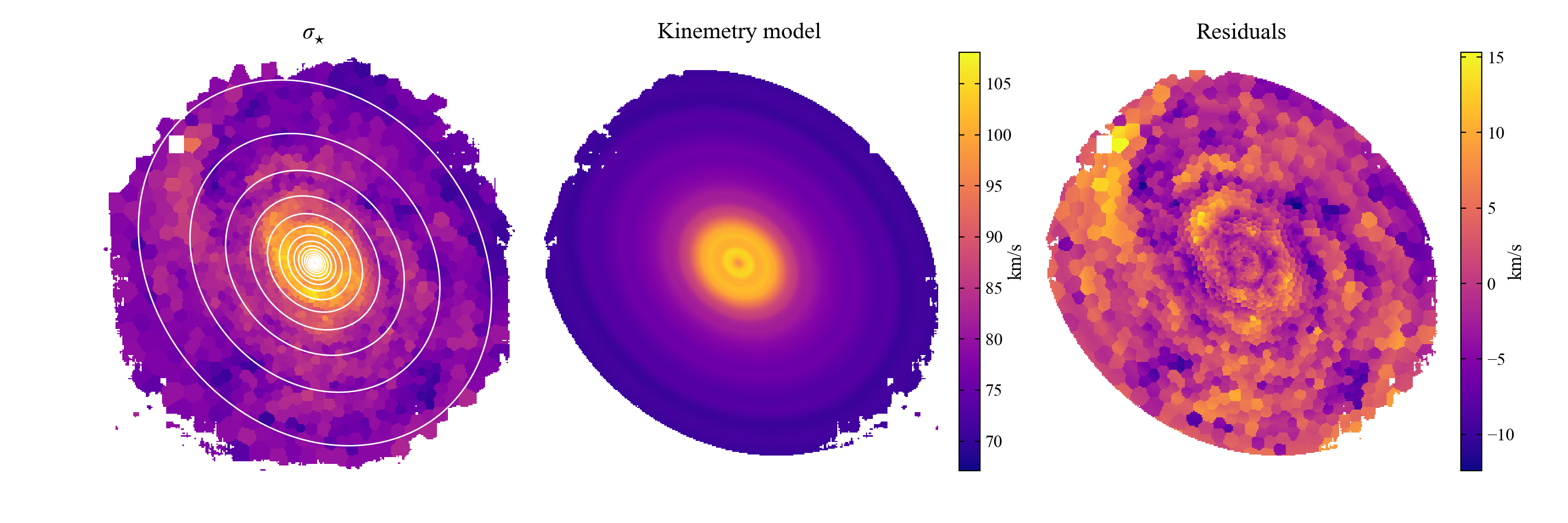}
\caption{Modelling and subtraction of the underlying \sigmastar\ substructure with \kinemetry.}
\label{app6}
\end{figure}

\begin{figure}[h!]
\centering
\includegraphics[width=1\linewidth]{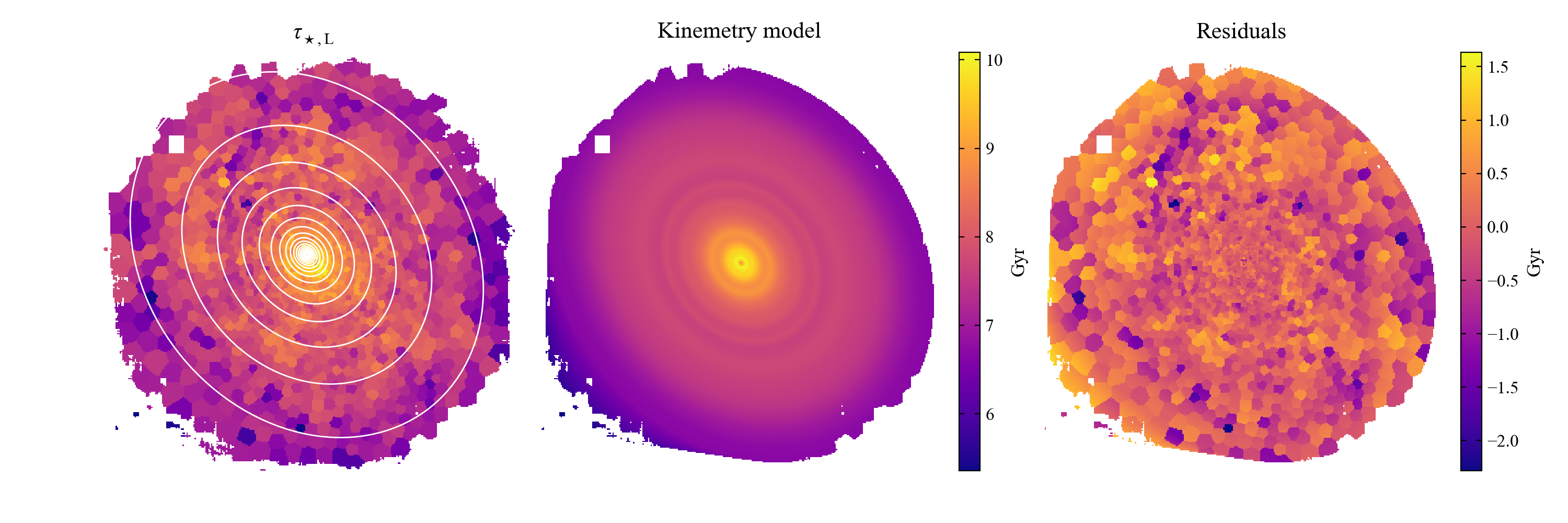}
\caption{Modelling and subtraction of the underlying $\tau_{\star,\rm L}$ substructure assessed by \fado\ with \kinemetry.}
\label{app7}
\end{figure}

\begin{figure}[h!]
\centering
\includegraphics[width=1\linewidth]{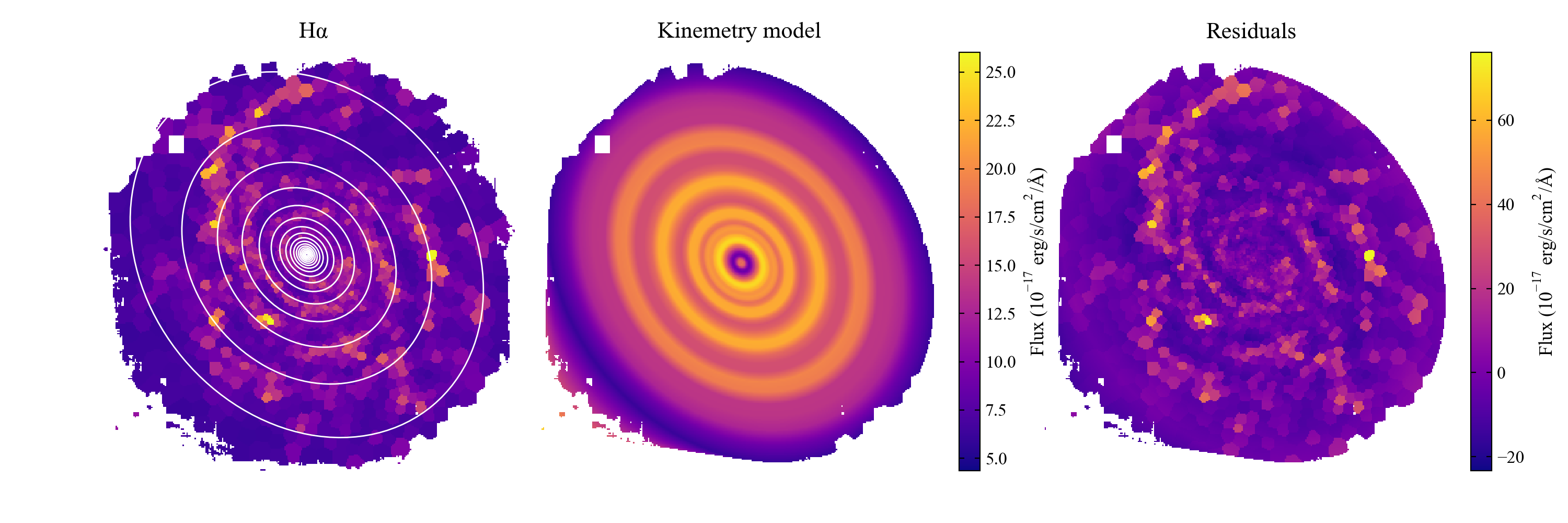}
\caption{Modelling and subtraction of the underlying H$\alpha$ substructure with \kinemetry.}
\label{app8}
\end{figure}

\end{appendices}

\end{document}